\documentclass{article}
\usepackage[utf8]{inputenc}
\usepackage{authblk}

\title{\textbf{Generalized potential for apparent forces: the Coriolis effect}}
\author{Elmo Benedetto$^{a}$, Ivana Bochicchio$^{b}$, Christian Corda$^{c}$,
Fabiano Feleppa$^{d}$,  Ettore Laserra$^{b}$ }
\date{}

\usepackage{graphicx}
\usepackage{amsmath}
\usepackage{bm}
\newtheorem{theorem}{Theorem}[section]
\newtheorem{remark}{Remark}[section]
\begin{document}
\maketitle

\begin{center}
$^{a}$ Department of Computer Science, University of Salerno, Via Giovanni Paolo II, 132, 84084 Fisciano (Sa), Italy
\\
$^{b}$ Department of Civil Engineering, University of Salerno, Via Giovanni Paolo II, 132, 84084 Fisciano (Sa), Italy
\\
$^{c}$ International Institute for Applicable Mathematics and Information Sciences, B. M. Birla Science Centre, Adarshnagar, Hyderabad 500063, India, and Department of Physics, Faculty of Science, Istanbul University, Istanbul, 34134, Turkey
\\
$^{d}$ Department of Physics, University of Trieste, via Valerio 2, 34127 Trieste, Italy
\end{center}

\begin{abstract}

\noindent It is well known, from Newtonian physics, that apparent forces appear when the motion of masses is described by using a non-inertial frame of reference. The generalized potential of such forces is rigorously analyzed focusing on their mathematical aspects.

\end{abstract}

\section{Introduction}

In the framework of classical mechanics, the configuration $P=(P_1,..,P_N)$ of a system with $n$ degree of freedom can described through the parametrization $P(t,q)$, where $t$ is the time and $q=(q_1, ..., q_n)$ is the Lagrangian parameter. The well known Eulero--Lagrange equation of motion is written as
\begin{equation} \label{lagrange}
\frac{d}{dt} \left( \frac{\partial T}{\partial \dot{q_i}}   \right) -\frac{\partial T}{\partial {q_i}} = Q_i ,
\end{equation}
where $T=T(t, q, \dot{q})$ is the kinetic energy, $\dot{q_i}= \frac{d q_i}{dt}$ are the generalized velocities and $Q_i$ the generalized components of the forces acting on the system defined as
\begin{equation} \label{Lagrange}
Q_i = \sum_{j=1}^N{\textbf{F}_j} \cdot \frac{\partial P_j}{\partial q_i}, \quad i=1,...,n .
\end{equation}
Now, for conservative system, it is possible to introduce the potential $U=U(q)$ such that 
\begin{equation} \label{generalized}
Q_i = \sum_{j=1}^N{\textbf{F}_j} \cdot \frac{\partial P_j}{\partial q_i}= \frac{\partial U }{\partial q_i}(q), \quad i=1,...,n ,
\end{equation}
where $\textbf{F}_j(t, P, \dot{P})$ are the total conservative forces applied to each point $P_i$. Hence, one can introduce the Lagrangian $L=T+U$ and consider the equation of motion in the equivalent form
\begin{equation} \label{L1}
\frac{d}{dt} \left( \frac{\partial L}{\partial \dot{q_i}}   \right) -\frac{\partial L}{\partial {q_i}} = 0 \quad i=1,...,n.
\end{equation}

Suppose now we have a generalized force that can be written in terms of a velocity-dependent
potential $U(q, \dot{q},t)$ as
\begin{equation} \label{potenzialiGen}
Q_{i} = \frac{\partial U}{\partial q_{i}} - \frac{d}{dt}\left(\frac{\partial U}{\partial \dot{q}_{i}}\right).
\end{equation}
If this is the case, substituting \eqref{potenzialiGen} into \eqref{lagrange}, we can conclude that the Euler-Lagrange equation still holds in the form \eqref{L1} for a Lagrangian function $L$ that can defined once more as $L = T+U$. The potential $U$ may be called a “generalized potential” or “velocity-dependent potential”. It is not a potential in the conventional sense because it depends on more than just the particle position and it cannot be calculated from a line integral of the generalized force. Despite this fact, the interesting aspect of such a formalism lies in the fact that it permits once more the use of a Lagrangian and the Euler-Lagrange equation and hence to re-obtain, in a generalized context, all the consequent properties. In addition, from a velocity-dependent potential can be derived interesting forces, such as the Lorentz force of a magnetic field $\textbf{B}$ acting on a moving charged particle 
\begin{equation} \label{lorentz}
\mathbf{F} = e (\mathbf{E} + \mathbf{v} \times \mathbf{B}),
\end{equation}
where $e$ is the particle charge, $\mathbf{v}$ the particle velocity, $\mathbf{E}$ the electric field. In particular, we will show how forces deriving from a generalized potential can be written in the form of \eqref{lorentz}, giving the suitable definition of the two vector fields $\textbf{E}$ and $\textbf{\textbf{B}}$. 
Following this line, the paper is so organized: In Section 2 we recall the expression of apparent forces in non-inertial frames. The generalized potential is analyzed in Section 3 and evaluated in Section 4 in case of apparent forces. 
%
%
\\
Finally, we (wish to) point out that the level of the paper is educational one, and hence, we restrict ourselves to examples from classical mechanics. Let's remark that this methods could be extended to the effects of relativistic mechanics

\section{Apparent forces in non-inertial frames}
Consider a free material point $(P, m)$, not subjected to effective forces, in absolute motion with respect to an inertial frame of reference $T_{\Omega} \equiv \Omega \xi \eta \zeta$ and in relative motion with respect to a non-inertial reference frame $T_{O} \equiv Oxyz$, moving in any rigid translational motion with respect to $T_{\Omega}$. If so, the principle of relatives motions is valid and we have
\begin{equation} 	\label{abs.vel}
\mathbf{v}^{(a)} = \mathbf{v} + \mathbf{v}_{\tau},
\end{equation}
where $\mathbf{v}^{(a)}$, $\mathbf{v}$ and $\mathbf{v}_{\tau}$ are respectively absolute, relative and translational velocity. In particular, if $\bm{\omega}_{\tau}$ is the angular velocity vector, the translational velocity is given by the well-known fundamental formula of rigid kinematics 
\begin{equation} \label{foundamentalEquation}
\mathbf{v}_{\tau} = \mathbf{v}_{O} + \bm{\omega}_{\tau} \times (P-O),
\end{equation}
where $\mathbf{v}_{O}$ is the velocity of the origin of the non-inertial reference frame $T_{O}$.
In its relative motion, the material point is subjected to the apparent forces
\begin{equation} \label{forza}
\mathbf{F} = -m\mathbf{a}_{\tau} - m\mathbf{a}_{c},
\end{equation}
where
\begin{equation} \label{trascinamento}
\mathbf{a}_{\tau} = \mathbf{a}_{O} + \dot{\bm{\omega}}_{\tau} \times (P-O) - \bm{\omega}_{\tau} \times (P-O) \times \bm{\omega}_{\tau}
\end{equation}
is the translational acceleration, and 
\begin{equation} \label{coriolis}
\mathbf{a}_{c} = 2\bm{\omega}_{\tau} \times \mathbf{v}
\end{equation}
is the Coriolis acceleration. Therefore
\begin{equation}
\mathbf{F}_{\tau} = -m\mathbf{a}_{\tau} = -m[\mathbf{a}_{O} + \dot{\bm{\omega}}_{\tau} \times (P-O) - \bm{\omega}_{\tau} \times (P-O) \times \bm{\omega}_{\tau}]    
\end{equation}
is the translational force (sometimes known as ‘dragging force’) and
\begin{equation}
\mathbf{F}_{c} = -m\mathbf{a}_{c} = -2m\bm{\omega}_{\tau} \times \mathbf{v}    
\end{equation}
is the Coriolis force. The sum of $\mathbf{F}_{\tau}$ and $\mathbf{F}_{c}$ provides the most general inertial force acting in the non-inertial reference
frame. Let us derive a generalized potential for such a force.The absolute motion of the point with respect to the inertial frame rests on Lagrange’s three scalar equations, that, in this case, since we use the three Cartesian coordinates as Lagrangian coordinates, may be summarized in the following vector equation:
\begin{equation}
\frac{d}{dt}\frac{\partial \mathbf{\mathcal{L}}^{(a)}}{\partial \mathbf{v}^{(a)}} - \frac{\partial \mathbf{\mathcal{L}}^{(a)}}{\partial P} = 0,
\end{equation}
where, by definition, we have
\begin{equation}
\left\{
\begin{array}{ll}
\frac{\partial}{\partial \mathbf{v}^{(a)}} = \mathbf{i}\frac{\partial}{\partial \dot{\xi}} + \mathbf{j}\frac{\partial}{\partial \dot{\eta}} + \mathbf{k}\frac{\partial}{\partial \dot{\zeta}}, & \\ & \\
\frac{\partial}{\partial P} = \mathbf{i}\frac{\partial}{\partial \xi} + \mathbf{j}\frac{\partial}{\partial \eta} + \mathbf{k}\frac{\partial}{\partial \zeta}. &
\end{array}%
\right.  \label{mp}
\end{equation}
Being the material point, in its absolute motion, not subjected to effective forces, the absolute Lagrangian $\mathcal{L}^{(a)}$ coincides with the absolute kinetic energy, so
\begin{equation}
\mathbf{\mathcal{L}}^{(a)} = \frac{m}{2}\mathbf{v}^{2}_{(a)} = T^{(a)}.
\end{equation}
To pass to the Lagrangian $\mathcal{L}$ in relative motion, we only have to substitute to the absolute velocity the expression given by \eqref{abs.vel}, so we obtain
\begin{equation} \label{lagrangian}
\mathcal{L} = \frac{m}{2}(\mathbf{v} + \mathbf{v}_{\tau})^{2} = \frac{m}{2}\mathbf{v}^{2} + m\left[\mathbf{v} \cdot \mathbf{v}_{\tau} + \frac{1}{2}\mathbf{v}^{2}_{\tau} \right],
\end{equation}
where $T = \frac{m}{2}\mathbf{v}^{2}$ is the apparent kinetic energy. Afterwards we will prove that the terms in square brackets as the second side of  \eqref{lagrangian} represent the generalized potential of apparent forces, not considering the sign and the mass $m$.
\begin{remark}
Following \eqref{Lagrange} we can obtain the generalized components $Q^D$ and $Q^{Cor}$ of the
dragging and Coriolis forces, respectively
\begin{equation} \label{generalDragging}
Q^D_h = -m[\mathbf{a}_{O} + \dot{\bm{\omega}}_{\tau} \times (P-O) + \bm{\omega}_{\tau} \times (P-O) \times \bm{\omega}_{\tau}] \cdot \frac{\partial P}{\partial q_h}
\end{equation}

\begin{equation} \label{generalCoriolis}
Q^{Cor}_h = -2m\bm{\omega}_{\tau} \times \mathbf{v}  \cdot \frac{\partial P}{\partial q_h}
\end{equation}

\end{remark}

\section{Deriving forces from a generalized potential}
In the physical space, the most general force resulting from a generalized potential can be expressed by the following classical theorem \cite{3}:
\begin{theorem} \label{teorema_potenziali}
If a force is of the type 
\begin{equation} \label{potentialforce}
\mathbf{F} = \mathbf{E} + \mathbf{v} \times \mathbf{B},
\end{equation}
where $\mathbf{v}$ is the velocity of the material point on which the force is exerted, than the two vectorial fields (eventually depending on time) $\mathbf{E}$ and $\mathbf{B}$ must satisfy
\begin{equation}
\left\{
\begin{array}{ll}
\nabla \cdot \mathbf{B} = 0, & \\ & \\
\nabla \times\textbf{ E} + \frac{\partial \mathbf{B}}{\partial t} = 0. &
\end{array}%
\right.  \label{mp}
\end{equation}
Moreover, a corresponding generalized potential is
\begin{equation}\label{eq:16}
U(P,\mathbf{v},t) = \phi(P,t) + \mathbf{A}(P,t) \cdot \mathbf{v},
\end{equation}
where $\phi(P,t)$ is a so-called scalar potential and $\mathbf{A}(P,t)$ is a so-called vector potential. In addition, the connection between force and generalized potential is provided by the two equations
\begin{equation}
\left\{
\begin{array}{ll}
\mathbf{B} = \nabla \times \mathbf{A}, & \\ & \\
\textbf{E} = \nabla \phi - \frac{\partial \mathbf{A}}{\partial t}. &
\end{array}%
\right.  \label{mp2}
\end{equation}
\end{theorem}
This is the case, for example, of the Lorentz electromagnetic force acting on a free point particle of charge $e$, position $P$ and velocity $\mathbf{v}$ in the presence of an electric field ${\textbf{E}}(t, P)$ and a magnetic induction ${\textbf{B}}(t, P)$:
\begin{equation}
\mathbf{F} =e  \mathbf{E} + e \mathbf{v} \times \mathbf{B}.
\end{equation}
This is a force of type \eqref{potentialforce} and the corresponding equations \eqref{mp} represent the first couple of the fundamental Maxwell equations. These equations are deduced as necessary conditions for the existence of a potential, without other physical considerations.
\section{Generalized potential of apparent forces}
This section is devoted to explicit the connection given in Equation \eqref{mp2} between the force of type \eqref{potentialforce} and generalized potential. 
Precisely, starting from a force of type \eqref{potentialforce} and following Theorem \ref{teorema_potenziali}, we known that it admits a generalized potential of type expressed in \eqref{eq:16}, i.e.
\begin{equation} \label{potenziale}
U(P,\mathbf{v},t) = \phi(P,t) + \mathbf{A}(P,t) \cdot \mathbf{v}. 
\end{equation}
Now, imposing
\begin{equation} \label{campovettoriale}
\mathbf{A}(P,t) = m \mathbf{v}_{\tau} = m [\mathbf{v}_{O} + \bm{\omega}_{\tau} \times (P - O)]
\end{equation}
as the vector potential and
\begin{equation} \label{camposcalare}
\phi(P,t) =  \frac{\mathbf{A}^{2}}{2} = m \frac{\mathbf{v}^{2}_{\tau}}{2}    
\end{equation}
as the ordinary scalar potential, we are interested to recover the explicit expression of the two vector fields $\mathbf{E}$ and $\mathbf{B}$.

\begin{remark} \label{remark}
It could be directly done by making explicit the terms appearing in the Lagrangian equations, but we prefer to use a different strategy based on theorem appearing in \cite[p. 64]{1}. Precisely, we observe that the two arguments $P$ and $t$ of the vector field $\mathbf{A}(P,t)$ must be considered as independent from each other. Furthermore, once the motion of the non-inertial reference frame $T_{O}$ with respect to the fixed (inertial) one $T_{\Omega}$, namely the translational motion, is assigned, the origin $O$, the velocity $\mathbf{v}_{O}$ and the angular velocity $\bm{\omega}_{\tau}$, become three known functions of time, $O = O(t)$, $\mathbf{v}_{O} = \mathbf{v}_{O}(t)$, $\bm{\omega}_{\tau} = \bm{\omega}_{\tau}(t)$ (see \cite[p. 70]{3}). 
\end{remark}
Let's consider \eqref{mp2}$_1$, that gives back (see for example \cite[Eq. (2.3.36) p. 293]{4})
\begin{equation}
\mathbf{B} = \nabla \times \mathbf{A} = m \nabla \times \mathbf{v}_{\tau} = 2 m \bm{\omega}_{\tau},
\end{equation}
so, by using also \eqref{coriolis}, we can write 
\begin{equation} \label{forza1}
\mathbf{F}  = \mathbf{E} + \mathbf{v} \times \mathbf{B}= \mathbf{E} - 2m \bm{\omega}_{\tau} \times \mathbf{v} = \mathbf{E} - m \mathbf{a}_{c}.
\end{equation}
To calculate the field $\mathbf{E}$ we use \eqref{mp2}$_2$, that gives 
\begin{equation} \label{field2}
\mathbf{E} = \nabla \phi(P,t) - \frac{\partial \mathbf{A}}{\partial t} = \frac{1}{2 m}\nabla \mathbf{A}^{2} - \frac{\partial \mathbf{A}}{\partial t}.
\end{equation}
We start to calculate the first term of the right hand side,
\begin{equation}
\frac{1}{2 m}\nabla \mathbf{A}^{2} = \frac{m}{2}\nabla \mathbf{v}^{2}_{\tau} = \frac{m}{2}\nabla(\mathbf{v}^{2}_{O} + 2\mathbf{v}_{O} \cdot \bm{\omega}_{\tau} \times (P - O) + [\bm{\omega}_{\tau} \times (P - O)]^{2}).
\end{equation}
Observing that $\nabla \mathbf{v}_{O}^2 = \textbf{0}$, we can write
\begin{equation}
\nabla [2\mathbf{v}_{O} \cdot \bm{\omega}_{\tau} \times (P - O)] = 2\nabla [\mathbf{v}_{O} \times \bm{\omega}_{\tau} \cdot (P - O)],
\end{equation}
and so
\begin{equation} \label{equaz}
\frac{1}{2}\nabla \mathbf{A}^{2} = 2m\nabla[\mathbf{v}_{O} \times \bm{\omega}_{\tau} \cdot (P - O)] + m\nabla[\bm{\omega}_{\tau} \times (P - O)]^{2}.
\end{equation}
Recalling that $(P-O)$ is a potential field hence $\nabla \times (P-O) = \bm0$ and using a well-known formula from the vector analysis (see for example \cite[p. 230]{2})
\begin{equation}
\nabla(\mathbf{A} \cdot \mathbf{B}) = (\mathbf{B} \cdot \nabla)\mathbf{A} + (\mathbf{A} \cdot \nabla)\mathbf{B} + \mathbf{B} \times \nabla \times \mathbf{A} + \mathbf{A} \times \nabla \times \mathbf{B},
\end{equation}
 we obtain
\begin{equation}
\nabla[\mathbf{v}_{O} \times \bm{\omega}_{\tau} \cdot (P - O)] = \mathbf{v}_{O} \times \bm{\omega}_{\tau}.
\end{equation}
By a substitution in \eqref{equaz} we can write
\begin{equation}
\frac{1}{2}\nabla \mathbf{A}^{2} = \frac{m}{2}\nabla[\bm{\omega}_{\tau} \times (P - O)]^{2} + m \mathbf{v}_{O} \times \bm{\omega}_{\tau}.
\end{equation}
If $P^{*}$ is the projection of $P$ on the instantaneous rotation axis related to the origin \textit{O}, that is the axis passing through $O$ and parallel to $\bm{\omega}_{\tau}$, we get
\begin{equation}
[\bm{\omega}_{\tau} \times (P - O)]^{2} = \bm{\omega_{\tau}}^{2}(P - P^{*})^{2}.
\end{equation}
If, at each fixed instant $t$, we introduce a system of cylindrical coordinates $r(=|P-P^{*}|),\theta, z$, having as its axis $z$ the instantaneous axis of rotation related to $O$, and define $\left \{ \hat{\mathbf{e}}_{1} = vers(P - P^{*}), \hat{\mathbf{e}}_{2}, \hat{\mathbf{e}}_{3} \right \}$ the associated basis (orthonormal in this case), the following expression results:
\begin{equation}
\bm{\omega}^{2}_{\tau}r^{2} = \bm{\omega}^{2}_{\tau}(P - P^{*})^{2},
\end{equation}
and, by recalling the gradient in cylindrical coordinates 
\begin{equation}
\nabla_{(r,\theta,z)} = \frac{\partial}{\partial r}\hat{\mathbf{e}}_{r} + \frac{1}{r}\frac{\partial}{\partial \theta}\hat{\mathbf{e}}_{\theta} + \frac{\partial}{\partial z}\hat{\mathbf{e}}_{z},
\end{equation}
we have
\begin{equation}
\nabla [\bm{\omega}_{\tau} \times (P - O)]^{2} = {\bm{\omega}}^{2}_{\tau}\frac{\partial r^{2}}{\partial r}\hat{\mathbf{e}}_{r} = 2\bm{\omega}^{2}_{\tau}r\hat{\mathbf{e}}_{r} = 2\bm{\omega}^{2}_{\tau}(P - P^{*}),
\end{equation}
that can be rewritten as
\begin{equation}
\frac{1}{2}\nabla [\bm{\omega}_{\tau} \times (P - O)]^{2} = \bm{\omega}_{\tau} \times (P - O) \times \bm{\omega}_{\tau}.
\end{equation}
Thus
\begin{equation} \label{eq2}
\frac{1}{2}\nabla \mathbf{A}^{2} = m \bm{\omega}_{\tau} \times (P - O) \times \bm{\omega}_{\tau} + m\mathbf{v}_{O} \times \bm{\omega}_{\tau}.
\end{equation}
Finally, following Remark \ref{remark},
\begin{equation} \label{eq3}
\frac{\partial \mathbf{A}}{\partial t} = m \frac{\partial}{\partial t}[\mathbf{v}_{O} + \bm{\omega}_{\tau} \times (P - O)] = m [\mathbf{a}_{O} + \dot{\bm{\omega}}_{\tau} \times (P-O) - \bm{\omega}_{\tau} \times \mathbf{v}_{O}].
\end{equation}
By substituting \eqref{eq2} and \eqref{eq3} into \eqref{field2} and using \eqref{trascinamento}, we obtain
\begin{equation} \label{E_finale}
\mathbf{E} = -m[ \mathbf{a}_{O} + \dot{\bm{\omega}}_{\tau} \times (P-O) - \bm{\omega}_{\tau} \times (P-O) \times \bm{\omega}_{\tau}] =m \mathbf{a}_{\tau} .
\end{equation}
Finally by \eqref{forza1} and \eqref{E_finale}, we conclude that
\begin{equation} \label{forza2}
\mathbf{F}  = \mathbf{E} + \mathbf{v} \times \mathbf{B}=  m \mathbf{a}_{\tau} - m \mathbf{a}_{c},
\end{equation}
that is exactly the translation forces \eqref{forza}.

\begin{remark}
Collecting \eqref{foundamentalEquation}, \eqref{potenziale}, \eqref{campovettoriale} and \eqref{camposcalare}, we can write the explicit expression of the generalized potential. Moreover, since $\frac{1}{2} m \mathbf{v}^2_O$ is certainly independent of $P$ and $\dot{P}$, it can be neglected and hence $U(P,\mathbf{v},t)$ can be regarded as a sum of three
contributions:
\begin{equation}
\label{eq:potenzialeFinale}
\begin{array}{lll}
U_1(\mathbf{v},t) = m \mathbf{v}_{O} \cdot \mathbf{v} \, ;
\\[1em]
U_2(P,\mathbf{v},t) = m \bm{\omega}_{\tau} \times (P - O) \cdot \mathbf{v} 
\\[1em]
U_3(P,t) =  m \mathbf{v}_O \cdot \bm{\omega}_{\tau} \times (P - O)+ \frac{1}{2}
m [\bm{\omega}_{\tau} \times (P - O)] ^2 
\end{array}
\end{equation}
Recalling \eqref{potenzialiGen}, it is possible to consider:
\begin{equation} \label{U1}
\frac{\partial U_1}{\partial q_{j}} - \frac{d}{dt}\left(\frac{\partial U_1}{\partial \dot{q}_{j}}\right) = - m\, \mathbf{a}_0 \cdot \frac{\partial P}{\partial q_{j} }
\end{equation}

\begin{equation} \label{U2}
\frac{\partial U_2}{\partial q_{j}} - \frac{d}{dt}\left(\frac{\partial U_2}{\partial \dot{q}_{j}}\right) = - 2 m\, \bm{\omega}_{\tau} \times \mathbf{v} \cdot \frac{\partial P}{\partial q_{j} }
- m \dot{\bm{\omega}}_{\tau} \times (P-O) \cdot \frac{\partial P}{\partial q_{j} }
- m \mathbf{v}_O \times \bm{\omega}_{\tau}  \cdot \frac{\partial P}{\partial q_{j} }
\end{equation}

\begin{equation} \label{U3}
\frac{\partial U_3}{\partial q_{j}} - \frac{d}{dt}\left(\frac{\partial U_3}{\partial \dot{q}_{j}}\right) = +  m\, \mathbf{v}_O \times \bm{\omega}_{\tau}  \cdot \frac{\partial P}{\partial q_{j} }
- m \bm{\omega}_{\tau} \times [\bm{\omega}_{\tau} \times (P-O)] \cdot \frac{\partial P}{\partial q_{j} } \,
\end{equation}
from which we obviously obtain \eqref{generalDragging} and \eqref{generalCoriolis}. From all these results we remark, as in \cite{siboni}, that,  a part of the case of uniformly rotating frames, we can't separate the contributions to the only time-dependent Coriolis force and to the only dragging force, respectively. In other words,  the generalized potential is a feature of the whole system of the general inertial forces and not separately to the the generalized components of each force.
\end{remark}

\section{Conclusion remarks}
In this paper we have reviewed the analogies between the electromagnetic force and the inertial ones. The generalized potential of Lorentz force is often studied in standard textbooks while the analogous potential of the Coriolis field is generally overlooked. We have highlighted this aspect describing the mathematical formalism of the inertial fields emphasizing the role of the Coriolis and Dragging forces.
%

\section{Acknowledgements}
The Authors thank the referees for useful comments.

\end{document}